# A Presentation Management System for Collaborative Meetings

K. Wrona
*DESY Hamburg, D-22603 Hamburg, Germany*

Electronic presentations are rapidly becoming the standard for meetings in large high energy physics collaborations. An attractive solution should combine a central repository of presentation files with easy uploading and downloading access over the wide area network. This allows all information to be instantly shared within the collaboration, including remote participants at meetings. An ideal system should allow operation of a whole session from a single PC, without time-consuming switchovers between personal notebooks, and preserve the information for summary talks etc. On the other hand, there should be no need for manual collection of documents, and submission of slides should be possible even seconds before the presentation.

We present a system developed and currently used by the ZEUS experiment at DESY. It provides a wide range of functions including registration of participants, automatic agenda preparation, presentation submission, display and archiving. A particular emphasis has been put on system security, with access to the system being made via a secure web interface. Different privilege levels are available within the system, according to the status of the user: chairman, speaker or other meeting participant. We also discuss our experience of using this system in a range of different contexts.

## 1. INTRODUCTION

A successful project in High Energy Physics demands a huge amount of information to be instantly shared among collaborators. Many working groups consist of scientists spread over the world, who spend most of their working time at their home institutes. Regular meetings are indispensable for a proper coordination of projects. Modern technology permits the organization of virtual conferences, which significantly improve the whole communication process. Apart from that, traditional meetings still play a big role in the scientific world. In both cases there is a necessity for maintaining a repository for presentation materials. Since manual collection of presentations involves a lot of effort and close cooperation between meeting organizers and speakers it is difficult to properly maintain such an archive. Very often slides become available only a long time after the meeting is over or in some cases they are not published and archived at all. This is particularly inconvenient for speakers, who must prepare a summary talk from a parallel session using materials from other presentations, as they usually do not have much time.

This paper presents the ZEMS system (ZEMS stands for **Z**EUS **E**lectronic **M**eeting Management **S**ystem), which fully automates the document collection procedure, and facilitates presentation and archiving. It has been developed within the ZEUS experiment at DESY.

## 2. KEY POINTS OF DESIGN

### 2.1. Simplicity

One of the main goals of this project was to create a tool which provides a simple interface to the end user. This is particularly important when the system is used for meetings with large audiences where any kind of problems and delays are not acceptable.

### 2.2. Availability

It should be possible to use the system from any place in the world where a network connection to the Internet is available.

### 2.3. Security

Security has been an essential consideration right from the start of the project. The whole transmission from client to web server is encrypted. Particularly, no cleartext password can be intercepted by possible intruders.

## 3. SYSTEM ARCHITECTURE

The ZEMS system is based on web technology. This allows everybody to use it without requiring any special software to be installed on the users desktop PC or notebook. The only necessary application is one of several widely available web browsers. The system is based on the Apache web server [1] with SSL support [2]. The PHP4 module [3] is used for the generation of the dynamic content of web pages, which act as an interface to the end user. All information typed into the web forms is passed to the system where the PHP module interprets the requests and processes them accordingly. If required the result is stored in a MySQL [4] database or in disk repository (fig. 1).

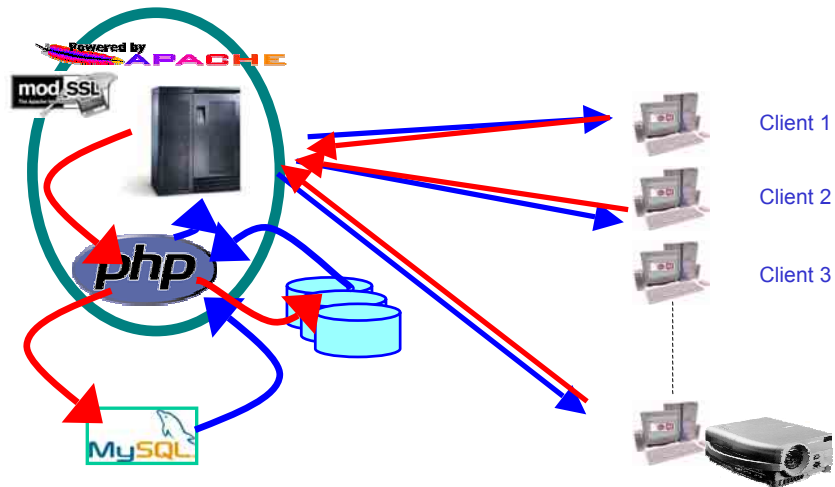

Figure 1. System architecture

## 4. THE ZEMS FUNCTIONALITY

The ZEMS system is an integrated tool which significantly improves the management of presentations for meetings. Its wide range of functions including, automatic agenda generation, presentation submission, display and archiving, will be described in the following.

### 4.1. Agenda generation

Automatic agenda generation reduces the amount of time invested by the manager during the preparation phase. There is no need to create a separate web site for each meeting or conference. The manager provides only the basic information required for a particular meeting, that is, defines sessions and presentations. In addition, for meetings with a high number of sessions, it is possible to delegate certain sub-tasks to other managers. For example, the detailed programme of each session can be prepared by its chairman. This reduces the amount of information which otherwise has to be exchanged between the chairman and the person responsible for the web site. Any modification can be made to the agenda on a very short timescale. The system provides a flexible mechanism for sub-task delegation on different levels. It also ensures that only managers with appropriate privileges can modify relevant information. In order to assign a speaker to a particular presentation each speaker must be registered within the ZEMS system. A personal account is created once and can be used for all meetings. The manager assigns speakers to presentations using a convenient interface. Limits can be imposed on the number of files each speaker is allowed to submit and their overall size.

### 4.2. Presentation upload

After the agenda is prepared the speaker may upload presentation files without any further assistance from meeting organizers. These files are immediately visible to all meeting participants. Files can be resubmitted and overwritten if needed. If the speaker does not wish to publish presentation in advance, the file can be hidden. The speakers may also sort the order of the files according to their preference. File can even be uploaded during an ongoing session, for example by participant attending the session with laptop and WLAN connection. This is a particularly convenient feature for summary speakers who often have little time for preparation. When using the system for less official meetings, for example videoconferences, it may be easily used for fast document exchange. The advantage of this approach is that all files remains in the repository and can be inspected after the meeting is over.

### 4.3. Presentation display

All meeting participants have easy access to all presentations via web pages. The whole meeting can be publicly available for viewing or protected by common password. The agenda is displayed on the web page with full information about each talk including links to corresponding slides and documents. The same access method is used in the auditorium for the actual presentation as on any desktop PC or notebook with access to the network.

### 4.4. Archiving

After the meeting is completed the manager can forbid any further changes. In this simple way a static archive is created.

## 5. HIERARCHY OF MEETINGS AND PRIVILEGES

A clear hierarchy has been established which allows for better management of the meetings. The whole system may be seen as one large container where different groups of meetings are defined according to subjects. Each group can contain a sizable number of meetings. A meeting consists of at least one session. For each session several presentations may be defined. Finally for each presentation the speaker may upload one or several files.

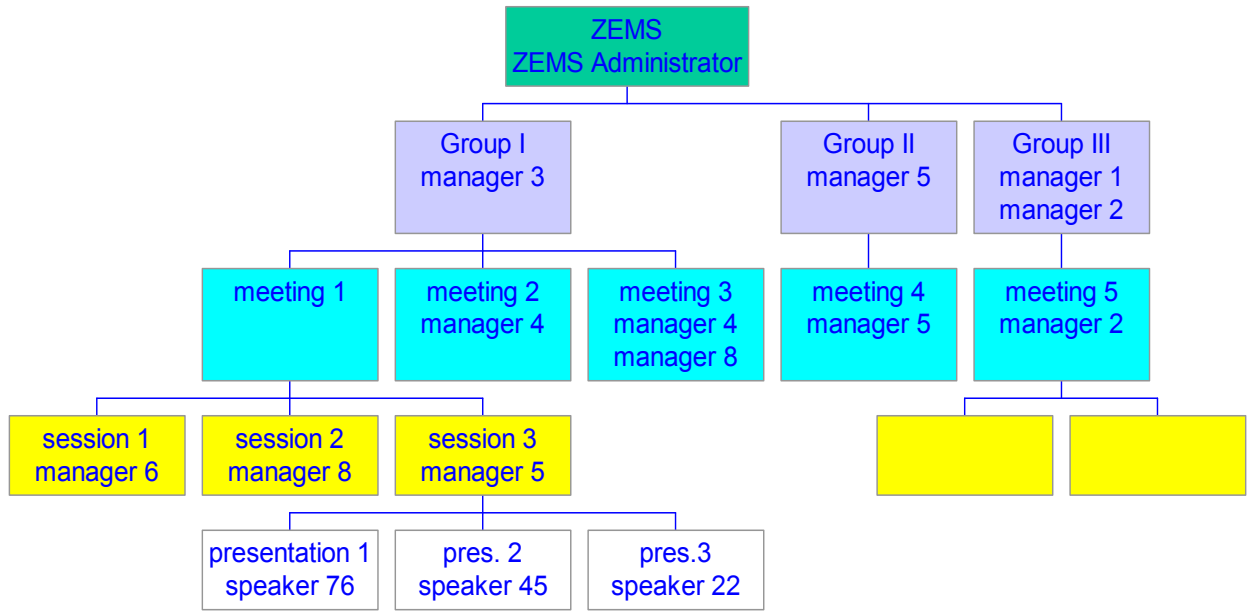

Figure 2: Hierarchy of meetings and privileges

Privileges may be granted to different managers at group, meeting and session level (fig.2). A speaker is assigned to each presentation.

## 6. INTERNAL STRUCTURE

The system consists of three modules. Each of them provides a set of functions according to the status of the user: manager, speaker or passive meeting participant. This separation allows for better control of privileges.

The manager part provides functions used for agenda generation, speakers account management and delegation of privileges. The speaker module provides the procedure for files submissions and management. Passive meeting participants use only the last module, which allows for agenda displaying and slide presentation. Only this module is used in the auditorium for presentation.

## 7. SECURITY ISSUES

Authentication is based on the users' registry stored in the MySQL database. Each user is identified by a unique ID and password. For each session, a secure connection is established between web browser and server using the SSL module for the Apache server. This way all information sent over the network is encrypted and cannot be read by possible intruders. In addition the password is sent only once and if authentication is successful a token is generated, which is valid for the remainder of the session. Based on user ID, appropriate privileges are obtained by the user. An access control list is also kept in the same database.

This kind of authentication and authorization has a significant impact on the way the file repository is maintained. The owner of the Apache process has read and write access to the disk. To prevent unauthorized access to the repository via a web browser the repository must be stored outside the scope of web server document tree. Hence, files cannot be accessed directly using URLs, because the built-in basic HTTP authentication mechanism is not used. The access is possible only via PHP functions. In addition the whole system must be installed on the dedicated web server with restricted user access.

## 8. IMPLEMENTATION DETAILS

The MySQL database is used to store all necessary information. For each level of the system (see fig. 2) a separate table is defined with fields corresponding to the attributes needed to describe it. For example, title, date, start and finish time, chairman and secretary characterize a session. Two tables contain all information about users of the system: speakers and managers. All passwords stored in these tables are encrypted. Another set of tables defines privileges for managers to a particular group, meeting or session.

The disk repository is organized according to the database structure. At the top level, the group directories are created with the names corresponding to the GROUP_ID field in the database table. The same convention is used for meetings, sessions and presentations subdirectories. This way files for each presentation are stored in separate directories and there is no need for speakers to use distinct file names for different presentations. The presentation file preserves an original name without the risk that it will be accidentally overwritten by other speakers.

## 9. WEB INTERFACE

The web interface to the ZEMS system has been implemented using the PHP4 scripting language. This technology allows the generation of dynamic web pages based on information stored in database. In particular, access to the MySQL database is well supported. Information supplied via the web forms is processed using scripts and appropriate updates to the database or web page are made.

Using only the web interface, managers can prepare the full agenda of a meeting (fig. 3). All previously supplied information may be modified or deleted. Announcements or comments can be placed on the meeting web page. Speakers are chosen using a drop-down menu. In addition the management of managers and speakers is done via the web interface.

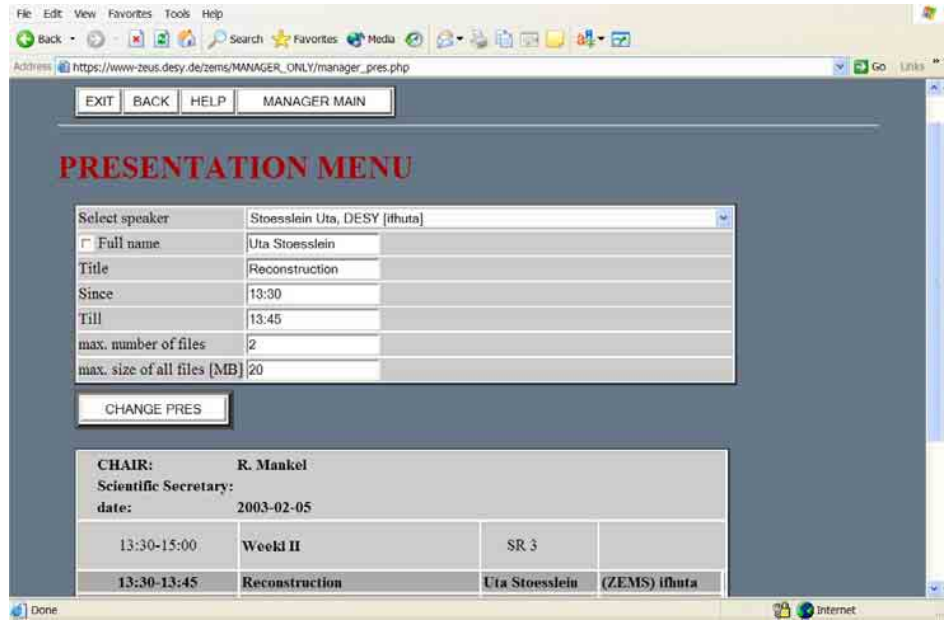

Figure 3. Example of the manager interface

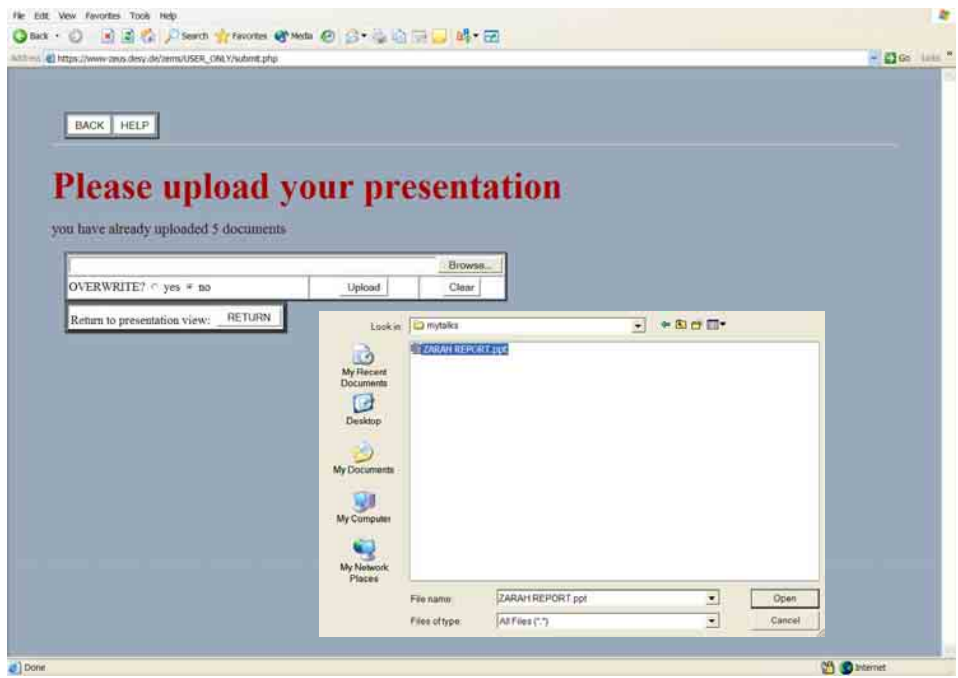

Figure 4. Presentation upload form

**PSN MONT009**

Speakers may easily upload presentation files via the web forms (fig. 4) and perform simple management of already uploaded files (fig. 5). These files can be ordered or hidden by clicking the buttons on the web page.

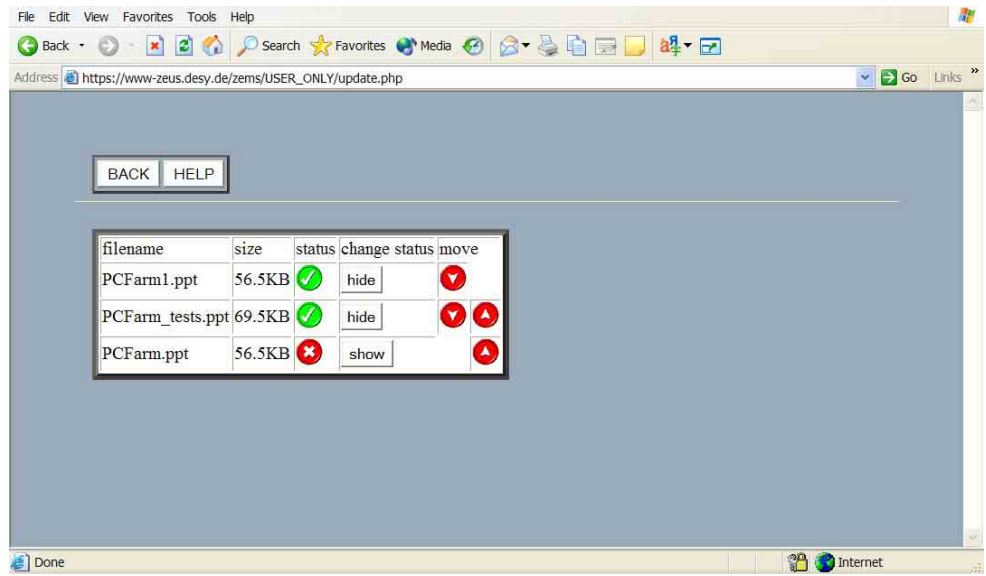

Figure 5. File management view

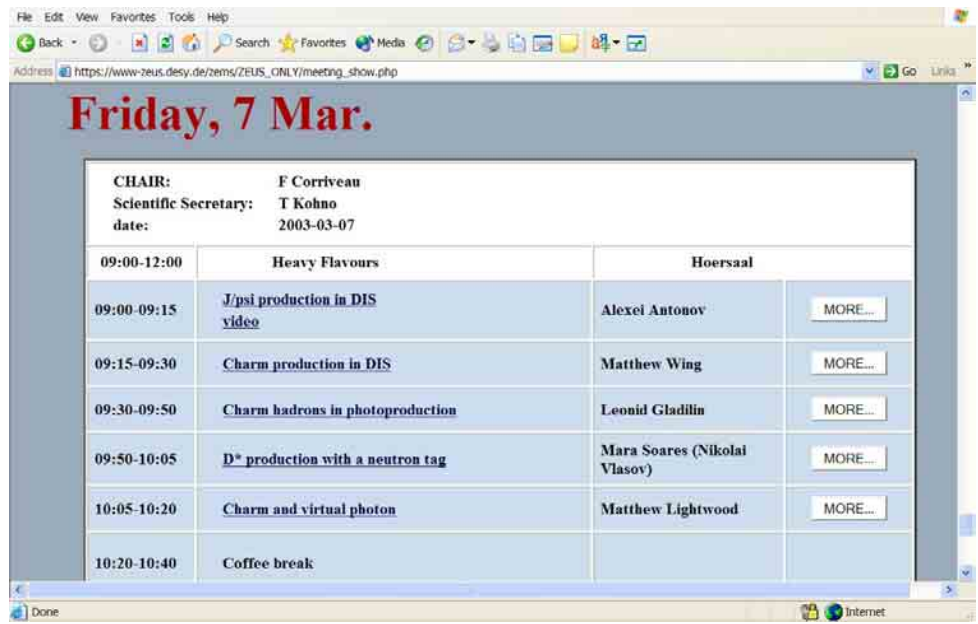

Figure 6. Meeting agenda view

A set of web pages exists that allows any meeting participant to easily find the appropriate presentation file. From the main page of the ZEMS system one follows the link to the required group, authenticates if necessary and then accesses a page for the meeting of interest (fig. 6).

The whole programme of the meeting is displayed in chronological order and presentation files are accessible via links. The agenda of the meeting and presentation slides can be displayed on the user desktop PC at the same time. In parallel, audio and video streams can be accessed if provided for the meeting (fig. 7).

Cascading Style Sheets [5] are used to define the color schemas. The user may choose one of them according to his preference. A special view for printing purpose is also available.

**PSN MONT009**

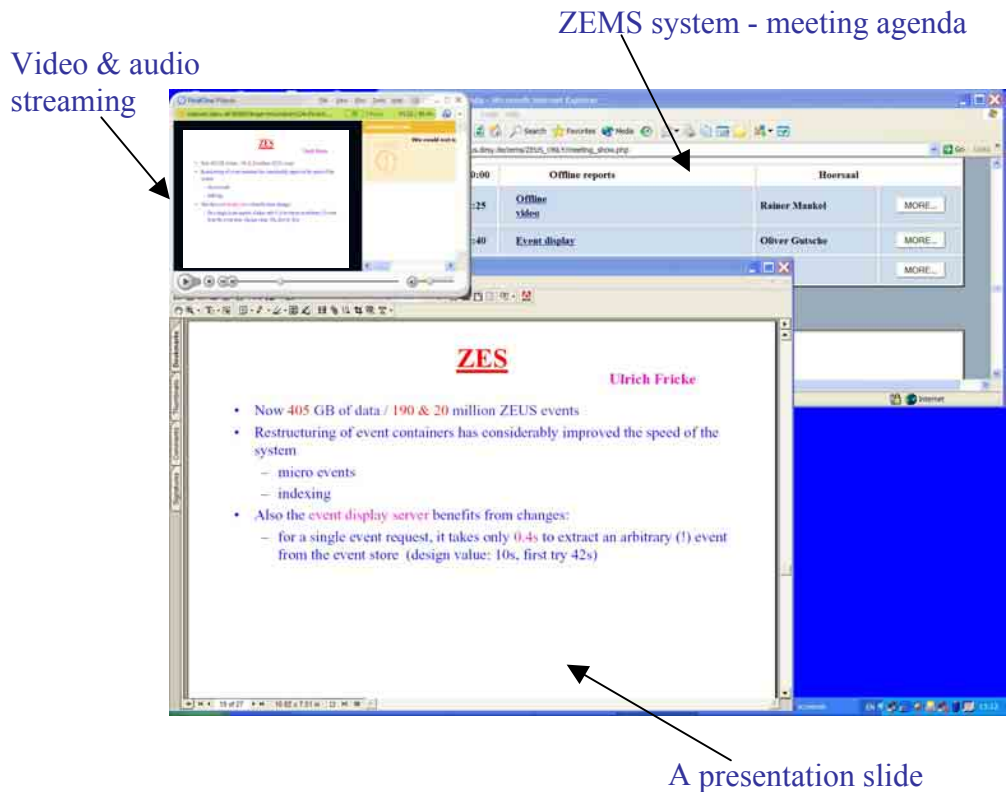

Figure 7. Example of the system usage by remote participant

## 10. EXPERIENCE

The system has obtained wide acceptance within the ZEUS collaboration. It is used for various kinds of meetings: collaboration meetings with a large number of participants as well as for small physics and technical working group meetings. It also supports videoconferences due to the fact that all uploaded materials are automatically archived. The first access to the system demands a speaker or manager to follow the registration procedure. Users who want to display submitted materials must only know the common password for a particular group if the access is protected.

The ZEMS system has been already used for more than 40 meetings. About 100 sessions have been defined with more than 500 presentations in total. There has not been a single failure of the system.

## 11. SUMMARY

The ZEMS system presented in this paper has been developed for presentations management at collaborative meetings. It provides an ergonomic interface for automatic agenda generation, easy and fast methods for presentations submission, retrieval and archiving. The wide availability to the system is assured by the web interface to the central repository with presentation files. The system supports secure connection and a password based authentication mechanism. This allows all information to be easily shared within the collaboration and prevents unauthorized access to the system.

## Acknowledgments

The author wants to thank Rainer Mankel for his support and many ideas he contributed, Radek Kaczorowski and Ingo Martens for their technical advice and help with system deployment.

## References

[1]   http://www.apache.org

[2]   http://www.modssl.org

[3]   http://www.php.net

[4]   http://www.mysql.com

[5]   http://www.w3.org/Style/CSS

**PSN MONT009**